\documentclass[12pt]{article}
\usepackage{amsmath,amssymb,bm}
\usepackage{amsthm}
\newtheorem{proposition}{Proposition}
\usepackage[T1]{fontenc}
\usepackage{lmodern}
\usepackage{booktabs}
\usepackage{array}
\usepackage{tabularx}
\usepackage{graphicx}
\usepackage{xcolor}
\usepackage{microtype}
\usepackage{enumitem}
\usepackage{natbib}
\usepackage{setspace}
\usepackage{titlesec}
\titlespacing*{\section}{0pt}{9pt plus 2pt minus 2pt}{4pt plus 1pt}
\titlespacing*{\subsection}{0pt}{7pt plus 2pt minus 2pt}{3pt plus 1pt}
\usepackage[font={small},skip=4pt]{caption}
\usepackage{tikz}
\usetikzlibrary{arrows.meta,positioning,calc,shapes.geometric}
\usepackage{xr-hyper}
\usepackage{hyperref}
\makeatletter
\newif\ifstoredxref
\@ifundefined{r@supp-app:priors}{\storedxreftrue\newlabel{supp-app:priors}{{S4}{7}{}{}{}}}{}
\@ifundefined{r@supp-fig:calibration}{\storedxreftrue\newlabel{supp-fig:calibration}{{S2}{22}{}{}{}}}{}
\@ifundefined{r@supp-fig:dag}{\storedxreftrue\newlabel{supp-fig:dag}{{S1}{6}{}{}{}}}{}
\@ifundefined{r@supp-sup:checks}{\storedxreftrue\newlabel{supp-sup:checks}{{S8}{20}{}{}{}}}{}
\@ifundefined{r@supp-sup:diagnostics}{\storedxreftrue\newlabel{supp-sup:diagnostics}{{S6}{12}{}{}{}}}{}
\@ifundefined{r@supp-sup:eiv-details}{\storedxreftrue\newlabel{supp-sup:eiv-details}{{S14}{34}{}{}{}}}{}
\@ifundefined{r@supp-sup:forecast}{\storedxreftrue\newlabel{supp-sup:forecast}{{S9}{22}{}{}{}}}{}
\@ifundefined{r@supp-sup:linkage}{\storedxreftrue\newlabel{supp-sup:linkage}{{S1}{2}{}{}{}}}{}
\@ifundefined{r@supp-sup:parameterization}{\storedxreftrue\newlabel{supp-sup:parameterization}{{S5.2}{9}{}{}{}}}{}
\@ifundefined{r@supp-sup:pooling}{\storedxreftrue\newlabel{supp-sup:pooling}{{S11}{27}{}{}{}}}{}
\@ifundefined{r@supp-sup:proofs}{\storedxreftrue\newlabel{supp-sup:proofs}{{S2}{2}{}{}{}}}{}
\@ifundefined{r@supp-sup:robustness}{\storedxreftrue\newlabel{supp-sup:robustness}{{S12}{28}{}{}{}}}{}
\@ifundefined{r@supp-sup:sim}{\storedxreftrue\newlabel{supp-sup:sim}{{S7}{14}{}{}{}}}{}
\@ifundefined{r@supp-sup:simfail}{\storedxreftrue\newlabel{supp-sup:simfail}{{S7.3}{15}{}{}{}}}{}
\@ifundefined{r@supp-tab:calibration-insample}{\storedxreftrue\newlabel{supp-tab:calibration-insample}{{S5}{21}{}{}{}}}{}
\@ifundefined{r@supp-tab:coverage}{\storedxreftrue\newlabel{supp-tab:coverage}{{S4}{19}{}{}{}}}{}
\@ifundefined{r@supp-tab:data-variables}{\storedxreftrue\newlabel{supp-tab:data-variables}{{S2}{3}{}{}{}}}{}
\@ifundefined{r@supp-tab:kappa}{\storedxreftrue\newlabel{supp-tab:kappa}{{S18}{34}{}{}{}}}{}
\@ifundefined{r@supp-tab:path-coefficients}{\storedxreftrue\newlabel{supp-tab:path-coefficients}{{S9}{24}{}{}{}}}{}
\@ifundefined{r@supp-tab:simulation}{\storedxreftrue\newlabel{supp-tab:simulation}{{S3}{17}{}{}{}}}{}
\@ifundefined{r@supp-tab:top-abilities}{\storedxreftrue\newlabel{supp-tab:top-abilities}{{S8}{23}{}{}{}}}{}
\@ifundefined{r@supp-tab:validation}{\storedxreftrue\newlabel{supp-tab:validation}{{S7}{23}{}{}{}}}{}
\makeatother
\ifstoredxref\AtBeginDocument{\PackageWarningNoLine{sync-xrefs}{%
  Could not read supplement.aux; using the numbers stored by sync_xrefs.py.\MessageBreak
  Recompile both documents and rerun that script before distributing this PDF}}\fi
\hypersetup{
  pdftitle={Beyond Match Results: Bayesian Mediation of Ability, Performance, and Winning in Professional Tennis},
  pdfkeywords={Bradley-Terry model; measurement error; pathway decomposition; sports analytics},
  colorlinks=true, linkcolor={blue}, citecolor={blue}, urlcolor={blue}}

\newcommand{\E}{\mathbb E}

\newcommand{\Normal}{\mathcal N}
\newcommand{\logit}{\operatorname{logit}}
\newcommand{\expit}{\operatorname{expit}}

\newcommand{\DE}{\operatorname{DE}}
\newcommand{\IE}{\operatorname{IE}}
\newcommand{\TE}{\operatorname{TE}}

\newcommand{\anon}{1}

\begin{document}

\def\spacingset#1{\renewcommand{\baselinestretch}%
{#1}\small\normalsize} \spacingset{1}

\if1\anon
{
  \title{\bf Beyond Match Results:\\ Bayesian Mediation of Ability, Performance, and Winning in Professional Tennis}
  \author{Guanyu Hu
    \hspace{.2cm}\\
    Department of Statistics and Probability, Michigan State University\\
    and \\
    Xiao Wang \\
    Department of Statistics, Purdue University}
  \date{}
  \maketitle
} \fi

\if0\anon
{
  \bigskip
  \bigskip
  \bigskip
  \begin{center}
    {\LARGE\bf Beyond Match Results:\\[4pt] Bayesian Mediation of Ability, Performance, and Winning in Professional Tennis}
  \end{center}
  \medskip
} \fi

\bigskip

\spacingset{1.2}
\begin{abstract}
Match results measure competitive ability, but they do not explain how ability
is expressed during play. Match statistics offer a natural mediator, however
performance is usually measured over the same match as the outcome: a
full-match statistic uses events that help determine the winner, while a
first-set statistic is earlier but rests on fewer service points and still
contributes to the result. Comparing the two therefore mixes timing and
precision. We develop a Bayesian mediation framework for this problem. Dynamic
abilities on hard, clay, and grass courts are estimated from previous seasons
only and are linked to both players' service performance, and a Bradley--Terry
model relates ability and performance to the match result. A control mediator
samples the first set's number of service points from the full match,
separating the measurement window from the point count. In 3,820 men's matches,
the hard-court mediated share rises from 0.52 with a first-set mediator to 0.95
with a full-match mediator; the point-count contrast accounts for 0.30 of this
gap and the window contrast for 0.13. The pattern replicates on the women's
tour. Simulations and sensitivity analyzes examine the decomposition and 
assumptions needed for a causal interpretation.
\end{abstract}

\noindent%
\vfill

\newpage
\spacingset{1.8} 

\section{Introduction}\label{sec:introduction}

The match results reveal differences in competitive ability, but they say little
about how those differences are expressed during play. A player may win
through a persistent advantage in ability because that advantage produces
better serving or returning, or because both pathways operate together.
Mediation analysis \citep{baron1986moderator,imai2010general} organizes such
questions, and paired-comparison models such as Bradley--Terry
\citep{bradley1952rank} have long measured player strength in tennis and other
sports \citep{glickman1999parameter,kovalchik2020elo,radicchi2011best}. Both
frameworks are well established, but combining them in tennis raises a
measurement problem that neither one addresses on its own.

Professional tennis is a useful setting for this question \citep{reid2010relationship,blackburn2013ranking,bozoki2016application}. The serve is the
most scrutinized part of the game: in Grand Slam play the server wins about
69--75\% of first-serve points and roughly 55\% of second-serve points, and
first-serve points are won about 69\% of the time on clay against 75\% on grass
and hard courts \citep{prietolage2023match}; point outcomes also carry
information beyond independent coin flips \citep{klaassen2001points}.
Match-level summaries of serve performance are widely used in broadcasts,
betting markets and player evaluation, but they are computed over the completed
match, so a full-match measure of service points won is built from events that
also help determine the winner. We call this \emph{outcome overlap}.

One apparent remedy is to shorten the measurement window, but restricting the
mediator to the first set changes more than the timing: fewer service points
enter the statistic, so it is measured less precisely, and classical
measurement error attenuates its coefficient and shrinks the estimated indirect
effect \citep{carroll2006measurement}. Comparing full-match and first-set
mediators therefore mixes a change in window with a change in precision, a
trade-off easy to overlook because within-match statistics feel like a direct
read on how a match was won. Diagnosing it requires varying the two separately,
which is what we develop below.

Existing methods stop short of such a diagnosis. The Bradley--Terry model
relates the probability that one player defeats another to the difference in
their latent abilities \citep{bradley1952rank}, in the paired-comparison
tradition initiated by \citet{zermelo1929berechnung} and developed for
computation and dynamic ability by
\citet{hunter2004mm,glickman1999parameter2,glickman1999parameter,cattelan2013dynamic};
the distinction between observed results and underlying strength matters in
sports because outcomes carry substantial contest-to-contest variation
\citep{lopez2018how}. In tennis, \citet{klaassen2001points} show that points
are neither independent nor identically distributed,
\citet{barnett2005combining} use service statistics for forecasting, and
\citet{ingram2019point} develops a hierarchical point-level model. Ranking and
Elo procedures remain strong predictive benchmarks
\citep{kovalchik2016search}, and \citet{dong2026dynamic} let player strength
depend on comparison-specific covariates in a Plackett--Luce model. None of
this work examines how a pre-match ability advantage is expressed through
performance measured during the match, or how the construction of that measure
affects the estimated pathway.

The mediation literature supplies the causal vocabulary but not the fix.
Direct and indirect pathways run from the regression formulation of
\citet{baron1986moderator} to the counterfactual definitions of
\citet{robins1992identifiability} and \citet{pearl2001direct}; identification,
estimation and sensitivity analysis are studied by
\citet{imai2010general,imai2010identification,vanderweele2015explanation},
with particular attention to unmeasured mediator--outcome confounding
\citep{vanderweele2010bias,cinelli2020making}. Timing is central to
time-varying exposures \citep{robins2000marginal}, and using information
collected during an outcome period is close to the logic of immortal time bias
\citep{suissa2008immortal}. Measurement error in a mediator is equally well
understood, but timing and precision are usually treated as separate concerns.
In tennis they change together, and that is the gap we address. Because latent
ability is not a manipulable treatment and the required mediator and outcome
assumptions cannot be verified from these data, we report direct and indirect
effects as model-based pathway summaries rather than unconditional causal
claims.

We make two methodological contributions. The first is a modular ability
model in which service performance cannot contaminate the ability assigned
before a match: a dynamic Bradley--Terry model pools player strength across
hard, clay and grass courts, ability entering a season-$t$ match is learned
only from results through season $t-1$, so the match under analysis never
contributes to its own pre-match estimate, and the performance and outcome
models are fitted conditional on draws from it
\citep{bayarri2009modularization,plummer2015cuts}, which prevents information
from flowing backward from within-match performance into pre-match ability.

The second is a point-count-matched design that separates what a shorter window
changes. The control mediator samples without replacement from all of a
player's service points, keeping the number served in the first set, so it
covers the full match at the first-set sample size. Comparing it with
the full-match statistic defines a point-count component; comparing it with
the first-set statistic defines the remaining window-construction component,
which may combine outcome overlap with genuine changes in performance across
sets. We therefore add a second outcome: first-set performance lies inside the
match outcome's window but entirely before the second-set outcome. These
contrasts distinguish point count from the remaining change in construction;
they do not separately identify overlap and performance change. Uncertainty in
constructing the control is propagated through a design-integrated posterior
mixture, analogous to multiple imputation \citep{rubin1987multiple}, and the
diagnostic makes the distinction explicit without claiming that either
contrast is automatically a causal effect.

The application contributes on its own terms. It asks not how to rank players
but how an ability advantage established before a match reaches the result
through service performance. Statistics such as service points won drive
broadcasts, betting lines and player evaluation, and the conclusions drawn from
them depend on two choices analysts usually make by default: the measurement
window and the number of points behind the statistic. The reliability analysis
extends the same logic to settings where only an aggregate statistic and its
denominator are observed, which covers most sports without point-level
records.

Section~\ref{sec:design} describes the linked match and point-level data.
Section~\ref{sec:model} introduces the surface-specific ability model, defines
the pathway estimands, states the identification conditions, describes
computation and reports the simulation evidence.
Section~\ref{sec:results} presents the ATP and WTA analyses with the
reliability analysis for settings without point-level data, and
Section~\ref{sec:discussion} concludes the paper with detailed discussions. Further simulations, computational
detail, diagnostics and posterior predictive checks are in the supplementary
material.

\section{Motivating data}
\label{sec:design}

\subsection{Data sources and study samples}

We combine tour-level match records with point-level charting data. The men's
and women's archives maintained by Jeff Sackmann provide main-draw singles
results, match dates, court surfaces, tournament rounds, rankings and
match-level service statistics \citep{sackmann_atp,sackmann_wta}; we retain
matches played on hard, clay and grass courts from 2015 through 2026. The
Match Charting Project supplies the point-level records needed to reconstruct
service performance over different portions of a match \citep{sackmann_mcp}.
Records are linked on normalized player names, match dates and surfaces, with
a small date discrepancy permitted only when the player pair and surface
agree.

The men's charting archive contains 6,575 records, of which 1,628 precede the
study period and 184 belong to the 2026 holdout season. Among the 4,763 records
from 2015 through 2025, 4,138 link to a tour-level match. Most of the remaining 625 could not link under any rule: 256 involve a player
outside the 485-player core used for ability estimation, 54 are qualifying-round
matches (20 of them in both groups) and 204 are Challenger matches, and the
main-draw tour files contain neither qualifying nor Challenger matches.
Section~\ref{supp-sup:linkage} of the supplementary material gives the
linkage rules and the disposition of every record. The women's replication is
constructed independently under the same definitions; the descriptive figures
in this section use the ATP samples only.

Table~\ref{tab:analysis-samples} summarizes the three ATP samples. The ability
sample is the broad set of tour matches used to estimate how player strength
changes across seasons and surfaces. The pathway sample is smaller because it
requires a link to the point-level archive, complete service information for
both competitors, a valid point-count-matched mediator and a prior ability
history. Matches from 2026 are reserved for forecast evaluation.

\begin{table}[htbp]
\centering
\setstretch{1}
\caption{Samples used in the primary ATP analysis.}
\label{tab:analysis-samples}
\small
\begin{tabularx}{\textwidth}{@{}l l r X@{}}
\toprule
Sample & Seasons & Size & Inclusion and use \\
\midrule
Ability
& 2015--2025
& 28,658 matches; 485 players
& Main-draw singles matches on hard, clay or grass, with both players
appearing in at least ten matches; used to estimate the dynamic
surface-specific ability model. \\

Pathway
& 2016--2025
& 3,820 matches
& Linked matches with complete first-set and full-match service records, a
valid point-count-matched mediator and prior ability information for both
players: 2,484 hard-court, 940 clay-court and 396 grass-court matches. \\

Holdout
& 2026
& 1,178 matches
& Matches reserved for forecast evaluation and excluded from estimation. \\
\bottomrule
\end{tabularx}
\end{table}

\begin{figure}[htbp]
\centering
\setstretch{1}

\includegraphics[width=\textwidth]{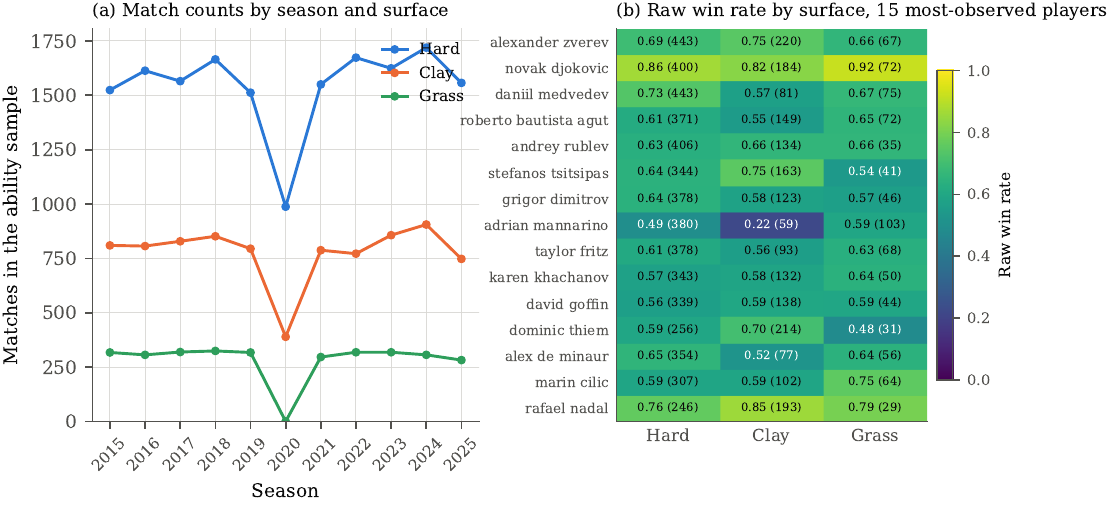}

\caption{Structure of the ATP ability sample. Panel (a) reports the number of
matches observed in each season on hard, clay, and grass courts. Panel (b)
reports raw surface-specific win rates for frequently observed players
represented on all three surfaces, with the corresponding match counts shown
in parentheses. The win rates are descriptive and do not adjust for opponent
strength.}
\label{fig:data-structure}
\end{figure}

Figure~\ref{fig:data-structure} describes the ability sample. The annual
match counts show how sharply the available information varies across seasons
and surfaces, and the player heatmap shows the same from the player side,
comparing raw surface-specific win rates for frequently observed players.
These rates do not adjust for opponent strength, but they make the central
point: a single rating can conceal large differences between hard, clay and
grass courts.

Before constructing the outcome we randomly assign the competitors in each
match to positions $a(m)$ and $b(m)$ using a fixed seed, removing the
winner-first ordering inherited from the source files; after randomization
player $a(m)$ wins 49.9\% of matches, as expected under an assignment
unrelated to the result. For player $i$ in match $m$, service performance over
a window $\mathcal W$ is
\begin{equation}
X_{im}(\mathcal W)
=
\frac{
\text{first-serve points won}
+
\text{second-serve points won}
}{
\text{service points played}
}.
\label{eq:service-stat}
\end{equation}
This is the proportion of all points won while serving, not the percentage of
first serves placed in play, and it can be reconstructed from the point-level
records over any window.

Table~\ref{supp-tab:data-variables} defines the outcomes, exposure and mediator
constructions used in the pathway analysis. The full-match mediator uses every
service point a player played; the first-set mediator uses only points from
the opening set; the point-count-matched control draws without replacement
from the complete match record while retaining that player's first-set
service-point count. All three come from the same point-level source and are
standardized within surface before fitting.

\begin{figure}[htbp]
\centering
\setstretch{1}

\includegraphics[width=\textwidth]{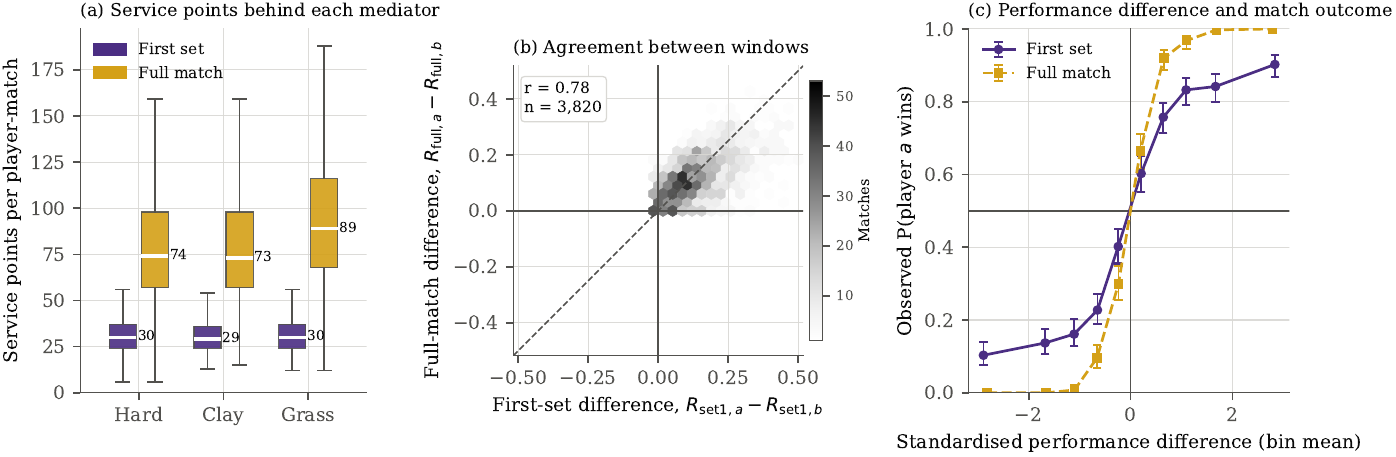}

\caption{Service-performance measurement in the ATP pathway sample. Panel (a)
compares the numbers of service points used to construct the first-set and
full-match mediators. Panel (b) compares the corresponding player performance
differences. Panel (c) reports the observed probability that player $a(m)$ wins
across binned values of the standardized performance difference. All panels use
the same matches included in the primary pathway analysis.}
\label{fig:mediator-windows}
\end{figure}

Figure~\ref{fig:mediator-windows} is built from the exact ATP matches in the
pathway analysis: its panels compare the service-point counts behind the
first-set and full-match mediators, the player differences from the two
windows, and the relation of each difference to the observed probability of
winning. The first-set statistic uses only about forty percent as many service
points, and while the two measures are positively associated, the full-match
difference aligns far more closely with the winner.

The second fact is why a mediation analysis of these data cannot proceed in the
usual way. The full-match mediator is assembled from the very points through
which the result is realized, so part of its association with winning is
mechanical rather than behavioral. The first-set statistic moves measurement
before most of the outcome, but discards more than half of the service points,
so comparing the two confounds a change in window with a change in precision,
and more matches cannot resolve a confound built into how the statistics are
defined. The point-count-matched control reproduces the loss of precision while
keeping the full-match window, so the two changes can be estimated separately.
It is sampled without replacement because sampling with replacement would add
variation from repeatedly drawing the same observed record.

\subsection{Empirical motivation}
\label{sec:questions}

Each feature of the ATP data rules out a simpler analysis. Players appear
repeatedly and their level changes over time, so a static rating is
insufficient. Their information differs sharply across surfaces, with long
hard-court histories and a short grass season, so three independent ratings
would be noisy where data are scarce while one pooled rating would suppress the
specialization the heatmap in Figure~\ref{fig:data-structure} suggests. The
ability model is therefore dynamic and hierarchical, letting overall strength
evolve while partial pooling carries information from well-observed surfaces to
sparse ones.

Service performance imposes a different requirement. A service-point percentage
reflects the server's quality and the returner's at once, so it is not a player
attribute in isolation, and the performance model conditions on the pre-match
abilities of both competitors. Its output enters the outcome model, where
Figure~\ref{fig:mediator-windows} shows a tight descriptive relationship
between the full-match performance difference and winning. That tightness is
the problem: it reflects outcome overlap as much as any behavioral pathway,
while shortening the window confounds timing with precision. The three mediator
constructions exist because the data offer no mediator that is at once early,
precise and free of overlap.

Temporal ordering creates one more constraint. An ability read off a model
fitted to the full study period would let current and future results leak into
the exposure assigned to earlier matches, so we estimate ability from expanding
windows of prior seasons and keep the performance model from feeding
information back. The design follows from what these data can and cannot
identify: repeated comparisons argue for a dynamic ability model, sparse
grass-court histories for partial pooling, opponent-dependent performance for a
two-competitor mediator, and the joint movement of timing and precision for the
point-count-matched control and the second-set comparison.

\section{Method}\label{sec:model}

\subsection{Dynamic surface-specific ability}
\label{sec:ability}

Let $i=1,\ldots,N$ index players, $m=1,\ldots,M$ index matches, and
$t$ index seasons. The surface of match $m$ is
$s_m\in\mathcal S=\{\mathrm{H},\mathrm{C},\mathrm{G}\}$, corresponding to hard,
clay, and grass courts, while $a(m)$ and $b(m)$ denote the two participating
players. We define $Y_m=1$ when player $a(m)$ wins and $Y_m=0$ otherwise.
The Bradley--Terry model represents this outcome through the difference
between the players' latent abilities \citep{bradley1952rank}. It belongs to
the paired-comparison tradition initiated by \citet{zermelo1929berechnung} and
has been extended to more general comparison structures, computational
procedures and abilities that evolve over time
\citep{hunter2004mm,glickman1999parameter2,cattelan2013dynamic}. The last
extension matters here, because the set of active players changes from season
to season and the information available on each surface is highly uneven.

A single rating is too restrictive here. Players develop persistent strengths
on particular surfaces, but an unrelated trajectory per surface would be
unstable, especially on grass, so we separate ability into an overall component
that evolves by season and a surface specialization shared across seasons:
\begin{equation}
\theta_{i,s,t}
=
\mu_{i,t}+\delta_{i,s},
\qquad
\mu_{i,t}
=
\mu_{i,t-1}+\omega_{it},
\qquad
\omega_{it}\sim\Normal(0,\sigma_\omega^2).
\label{eq:theta}
\end{equation}
Here $\mu_{i,t}$ is the player's competitive level in season $t$ and
$\delta_{i,s}$ the amount by which surface $s$ raises or lowers it. Unobserved
annual innovations are integrated out over an absence, so for consecutive
active seasons $t'<t$,
\[
\mu_{i,t}\mid\mu_{i,t'}
\sim
\Normal\!\left(
\mu_{i,t'},(t-t')\sigma_\omega^2
\right).
\]
This is the transition implied by the unconstrained annual increments, not an
assumption that ability is fixed during an absence. Because the zero-sum
restrictions below couple the players' coordinates, it is a building block of
the constrained Gaussian prior rather than the conditional distribution of one
constrained coordinate.

Conditional on the latent abilities, match results follow the
surface-specific Bradley--Terry model
\begin{equation}
Y_m\mid\bm\theta,\bm a
\sim
\operatorname{Bernoulli}(p_m),
\qquad
\logit(p_m)
=
a_{s_m}
\left(
\theta_{a(m),s_m,t_m}
-
\theta_{b(m),s_m,t_m}
\right),
\label{eq:bt}
\end{equation}
where $a_s>0$ is a surface-specific discrimination parameter. The ability
difference determines which player is favored; $a_s$ how sharply latent
separation translates into a difference in win probability. Letting it vary by
surface avoids forcing the same latent gap to carry the same predictive
consequence on every court.

The likelihood in \eqref{eq:bt} depends only on ability differences, and the
decomposition in \eqref{eq:theta} introduces an additional invariance between
overall ability and surface specialization. We remove these invariances by
imposing
\begin{equation}
\sum_{i\in\mathcal A_t}\mu_{i,t}=0
\quad\text{for every }t,
\qquad
\sum_{i=1}^N\delta_{i,s}=0
\quad\text{for every }s,
\qquad
\sum_{s\in\mathcal S}\delta_{i,s}=0
\quad\text{for every }i,
\label{eq:constraints}
\end{equation}
where $\mathcal A_t$ is the set of players active in season $t$. The first
restriction defines each season relative to its active player population; the
others separate general ability from surface specialization, since surface
effects average to zero across players and each player's across the three court
types. The restrictions are imposed through orthonormal Helmert bases: with
$H_t$ the basis for the players active in season $t$ and $H_N$ and $H_3$ the
player and surface bases,
\[
\bm\mu_t=H_t\bm z_t,
\qquad
\bm\delta
=
\sigma_\delta H_ND^{\mathrm{raw}}H_3^\top,
\qquad
\operatorname{vec}(D^{\mathrm{raw}})
\sim\Normal(\bm 0,I).
\]
This parameterization is one-to-one on the constrained space and avoids the
redundant direction left by centering unrestricted coordinates after sampling.
The noncentred representation of $\bm\delta$ also reduces posterior dependence
between the surface effects and their scale $\sigma_\delta$.

The zero-sum restriction changes the normalizing term of the random-walk prior,
which depends on $\sigma_\omega$ and is retained when that variance is
estimated (Section~\ref{supp-sup:parameterization} of the supplementary
material gives the expression). The scale is anchored through
$\mu_{i,t_i^{\min}}\sim\Normal(0,1)$, with $t_i^{\min}$ player $i$'s first
season: rescaling every ability and dividing $a_s$ by the same constant leaves
\eqref{eq:bt} unchanged, so $a_s$ is read relative to this reference scale.

The pathway analysis requires ability to be determined before the match under
study. A trajectory fitted to all seasons fails this, because the random walk
smooths an earlier state using later results. For every target season $T$ we
therefore refit \eqref{eq:theta} and \eqref{eq:bt} using only matches played
before $T$. Let $t_i^*(T)<T$ be the most recent season in that window in which
player $i$ was active and $g_i(T)=(T-1)-t_i^*(T)$. The exposure is the latent
state at the end of the historical window, season $T-1$, not an unobserved
season-$T$ state. Propagating the fitted transition model to that boundary
gives
\begin{equation}
\widetilde\theta^-_{i,s}(T)
=
\mu_{i,t_i^*(T)}
+\delta_{i,s}
+\sigma_\omega\sqrt{g_i(T)}\,Z_i,
\qquad
Z_i\sim\Normal(0,1).
\label{eq:theta-pre}
\end{equation}
The extra innovation term vanishes for a player observed in the preceding
season, whose posterior draw already carries that season's uncertainty, and
propagates the unobserved transitions through season $T-1$ for a player
returning after an absence; a one-step forecast of the season-$T$ state would
use $T-t_i^*(T)$ instead, which is not the exposure here. Players without
earlier match history are excluded rather than assigned an exposure determined
almost entirely by the prior.

Because each expanding-window fit has its own anchored scale, the abilities
passed to the pathway model are standardized within their window:
\begin{equation}
\theta^-_{i,s}(T)
=
\frac{\widetilde\theta^-_{i,s}(T)}{\varsigma_T},
\qquad
\varsigma_T
=
\operatorname{sd}_{i,s}
\left[
\E\left\{
\widetilde\theta^-_{i,s}(T)
\mid\mathcal D_{<T}
\right\}
\right],
\label{eq:theta-standardized}
\end{equation}
where the standard deviation is taken over players with prior history and the
three surfaces, so one unit of $\theta^-$ is one window-specific standard
deviation in the historical ability field. The eligible player population is
defined using the complete study period, so sample membership stays
retrospective even though no current or future outcome enters $\theta^-$. That
does not affect comparisons among mediator constructions, which share the same
ability draws, but it limits the standardized unit to the population the study
represents.

\subsection{Observed performance and match outcomes}
\label{sec:pathway-model}

The second stage describes how pre-match ability is expressed through observed
service performance and how both quantities are associated with the match
result. Let
$w\in\mathcal W=\{\mathrm{set1},\mathrm{ctrl},\mathrm{full}\}$ index the
mediator construction, and let $R^{(w)}_{im}$ denote the proportion of service
points won by player $i$ in match $m$. The three constructions use the first
set, a point-count-matched subsample of the full match, and the full match,
respectively. Because
the marginal distribution of service performance differs across surfaces, each
version is centered and scaled within surface, pooling the observations from
both players:
\begin{equation}
X^{(w)}_{im}
=
\frac{
R^{(w)}_{im}-\overline R^{(w)}_{s_m}
}{
S^{(w)}_{s_m}
}.
\label{eq:mediator-standardization}
\end{equation}
For a resampled control the standardization is repeated within each imputation,
using the control realization supplied to that fit. We suppress $w$ when
describing one construction, but all pathway parameters are
construction-specific, with $c_s^{(w)}$, $\lambda_s^{(w)}$ and $g_s^{(w)}$
estimated in separate fits: the three constructions are alternative
measurements of the mediator, not repeated observations governed by a common
coefficient.

Service performance depends on both players: a strong server wins more points
on serve, but fewer of them against a stronger opponent. We represent both
features jointly as
\begin{align}
X_{a(m)m}
&=
\alpha_{s_m}
+\beta^{\mathrm{own}}_{s_m}\theta^-_{a(m),s_m}
+\beta^{\mathrm{opp}}_{s_m}\theta^-_{b(m),s_m}
+\varepsilon_{am},
\label{eq:xa}\\
X_{b(m)m}
&=
\alpha_{s_m}
+\beta^{\mathrm{own}}_{s_m}\theta^-_{b(m),s_m}
+\beta^{\mathrm{opp}}_{s_m}\theta^-_{a(m),s_m}
+\varepsilon_{bm},
\label{eq:xb}\\
\begin{pmatrix}
\varepsilon_{am}\\
\varepsilon_{bm}
\end{pmatrix}
&\sim
\Normal\left[
\begin{pmatrix}0\\0\end{pmatrix},
\sigma_X^2
\begin{pmatrix}
1&\rho\\
\rho&1
\end{pmatrix}
\right].
\label{eq:xcov}
\end{align}
The restrictions $\beta^{\mathrm{own}}_s>0$ and $\beta^{\mathrm{opp}}_s<0$
orient the model so that own ability improves service performance and opponent
ability suppresses it. The correlation $\rho$ is unrestricted: match conditions
may move the two residuals together, while momentum or other within-match
dynamics may move them apart, and the bivariate distribution allows either
pattern rather than treating the two player-match observations as
conditionally independent.

Equations~\eqref{eq:xa}--\eqref{eq:xcov} are a working model for the
surface-standardized proportion; they do not use the service-point count as a
binomial denominator, which enters instead through the control mediator below.
Reliability and simulation analyses assess departures from the assumptions that
make that control comparable to the first-set measure.

The outcome layer extends the Bradley--Terry model by allowing realized service
performance to enter alongside the pre-match ability difference:
\begin{equation}
Y_m
\sim
\operatorname{Bernoulli}(p_m),
\qquad
\logit(p_m)
=
c_{s_m}\Delta\theta^-_m
+\lambda_{s_m}\Delta X_m,
\label{eq:outcome}
\end{equation}
where
\[
\Delta\theta^-_m
=
\theta^-_{a(m),s_m}-\theta^-_{b(m),s_m},
\qquad
\Delta X_m
=
X_{a(m)m}-X_{b(m)m}.
\]
The surface-specific coefficient $c_s>0$ measures the part of the ability
association remaining after conditioning on observed performance, and
$\lambda_s>0$ the association between the service-performance advantage and the
outcome at a fixed ability difference. Equation~\eqref{eq:outcome} is fitted
separately for each mediator construction and each outcome. With the match
winner, the first-set mediator occurs earlier but remains part of the
outcome-generating period; with the second-set winner, first-set performance is
measured entirely before the outcome window.

The positivity restrictions orient the decomposition and imply a mediated share
between zero and one. They are substantive rather than identification
conditions and exclude inconsistent mediation, so we also report an
unrestricted refit of the direct coefficient and read near-boundary shares
against it.

The two stages are kept modular, because a joint posterior would let the
within-match mediator and outcome revise the latent ability assigned to that
match, weakening the pre-match interpretation the expanding-window construction
establishes. With $\mathcal D_A$ the historical results used to estimate
ability, $\mathcal D_P$ the pathway data and $\bm\psi$ the performance and
outcome parameters, we target the cut distribution
\begin{equation}
p_{\mathrm{cut}}
\left(
\bm\psi,\bm\theta^-
\mid\mathcal D_A,\mathcal D_P
\right)
=
p\left(
\bm\psi\mid\bm\theta^-,\mathcal D_P
\right)
p\left(
\bm\theta^-\mid\mathcal D_A
\right),
\label{eq:cut-joint}
\end{equation}
whose marginal pathway distribution is
\begin{equation}
p_{\mathrm{cut}}(\bm\psi)
=
\int
p\left(
\bm\psi\mid\bm\theta^-,\mathcal D_P
\right)
p\left(
\bm\theta^-\mid\mathcal D_A
\right)
\,d\bm\theta^-.
\label{eq:cut}
\end{equation}
The first-stage posterior acts as an imputation distribution for pre-match
ability, and the second stage is refitted conditionally on draws from it
\citep{rubin1987multiple}, propagating ability uncertainty without feedback
from the pathway data. The result is a modular, or cut, posterior rather than
an ordinary joint posterior \citep{plummer2015cuts,jacob2017better}.

Contrasts among mediator constructions require a joint convention for the
otherwise separate fits. We use the comparison distribution
\begin{equation}
q_{\mathrm{cmp}}
\left(\bm\theta^-,\{\bm\psi^{(w)}\}_{w\in\mathcal W}\right)
=
p(\bm\theta^-\mid\mathcal D_A)
\prod_{w\in\mathcal W}
p\left(\bm\psi^{(w)}\mid\bm\theta^-,\mathcal D_P^{(w)}\right).
\label{eq:comparison}
\end{equation}
Figure~\ref{supp-fig:dag} of the supplementary material shows the resulting
temporal structure, with an unobserved match-level shock that may act on both
mediator and outcome. All constructions share the same draw of pre-match
ability, while their pathway parameters are conditionally independent given
that draw. This is a comparison distribution, not a generative joint posterior:
the same match outcome appears in each alternative analysis and is not
generated three times. Equation~\eqref{eq:comparison} states the dependence
used for posterior contrasts without suggesting that the data identify residual
cross-model covariances that were never estimated.

\subsection{Pathway estimands and identification}
\label{sec:estimands}

A pathway contrast must specify how the abilities of both competitors change.
We use a symmetric intervention that changes their difference while holding
their mean fixed:
\begin{equation}
\theta_a(d)=\frac{d}{2},
\qquad
\theta_b(d)=-\frac{d}{2},
\qquad
\Delta\theta(d)=d.
\label{eq:intervention}
\end{equation}
This avoids reading the contrast as a change in the general standard of the
match. Under \eqref{eq:xa} and \eqref{eq:xb}, the expected performance
difference responds to $d$ through
\begin{equation}
g_s
=
\frac{\partial}
{\partial d}
\E\{\Delta X_m(d)\mid s\}
=
\beta^{\mathrm{own}}_s-\beta^{\mathrm{opp}}_s.
\label{eq:mediator-gradient}
\end{equation}
Since $\beta^{\mathrm{own}}_s>0$ and $\beta^{\mathrm{opp}}_s<0$, the gradient
$g_s$ is positive: raising one player's ability while lowering the opponent's
widens the expected performance difference through both sides of the match.

For construction $w$ and the reference contrast $d_0=0$ and $d_1=1$, write
\[
\Delta X_m^{(w)}(d)=g_s^{(w)}d+e^{(w)},
\qquad
e^{(w)}\sim
\Normal\left[0,2\{\sigma_X^{(w)}\}^2\{1-\rho^{(w)}\}\right].
\]
The surface intercept drops out of the difference and the common part of the
two residuals cancels. Using the same realization of $e^{(w)}$ under both
ability levels, the direct, indirect and total pathway summaries on the
win-probability scale are
\begin{align}
\DE_s^{(w)}
&=
\E_{e^{(w)}}\left[
\expit\{c_s^{(w)}+\lambda_s^{(w)}e^{(w)}\}
-\expit\{\lambda_s^{(w)}e^{(w)}\}
\right],
\label{eq:de}\\
\IE_s^{(w)}
&=
\E_{e^{(w)}}\left[
\expit\!\left[c_s^{(w)}+\lambda_s^{(w)}\{g_s^{(w)}+e^{(w)}\}\right]
-\expit\{c_s^{(w)}+\lambda_s^{(w)}e^{(w)}\}
\right],
\label{eq:ie}\\
\TE_s^{(w)}
&=
\E_{e^{(w)}}\left[
\expit\!\left[c_s^{(w)}+\lambda_s^{(w)}\{g_s^{(w)}+e^{(w)}\}\right]
-\expit\{\lambda_s^{(w)}e^{(w)}\}
\right].
\label{eq:te}
\end{align}
The direct effect changes ability with the mediator held at its $d_0$
distribution; the indirect effect moves the mediator from its $d_0$ to its
$d_1$ distribution with the ability contrast held at $d_1$; the total effect
changes both. These definitions give
\[
\TE_s^{(w)}=\DE_s^{(w)}+\IE_s^{(w)},
\]
and the construction-specific mediated share
$S_s^{(w)}=\IE_s^{(w)}/\TE_s^{(w)}$, both evaluated within each posterior
draw. Numerical verification of the identity is a coding check, not evidence
for the model.

The decomposition reported in the application is now explicit. For each
surface, define
\begin{align}
G_s
&=S_s^{(\mathrm{full})}-S_s^{(\mathrm{set1})},\nonumber\\
P_s
&=S_s^{(\mathrm{full})}-S_s^{(\mathrm{ctrl})},
\qquad
W_s
=S_s^{(\mathrm{ctrl})}-S_s^{(\mathrm{set1})}.
\label{eq:share-decomposition}
\end{align}
Here $G_s$ is the conventional full-match versus first-set gap, $P_s$ the
point-count component and $W_s$ the remaining window-construction component,
with $G_s=P_s+W_s$ by definition. These are contrasts among three fitted
constructions for a centered pair separated by one standard deviation of
historical ability, not population average treatment effects; the labels do not
by themselves supply a causal interpretation.

The control separates the number of service points from the part of the match
over which they are observed. With
$\mathcal R_{im}=(W_{im1},\ldots,W_{imN_{im}})$ the service-point outcomes of
player $i$ in match $m$, write
\[
R^{\mathrm{full}}_{im}
=
\frac{1}{N_{im}}
\sum_{k=1}^{N_{im}}W_{imk}.
\]
If the player serves $n_{im}$ points in the first set, the control mediator is
the mean of $n_{im}$ entries drawn uniformly without replacement from
$\mathcal R_{im}$, independently of their values.

\begin{proposition}[Common target of the full and control mediators]
\label{prop:precision}
Conditional on the complete service record,
\[
\E\left(
R^{\mathrm{ctrl}}_{im}\mid\mathcal R_{im}
\right)
=
R^{\mathrm{full}}_{im}.
\]
Consequently, if
$\E(R^{\mathrm{full}}_{im}\mid\pi_{im})=\pi_{im}$ for a latent match-specific
service rate $\pi_{im}$, then
\[
\E(R^{\mathrm{ctrl}}_{im}\mid\pi_{im})=\pi_{im}.
\]
The full-match and control mediators therefore target the same performance
quantity. Their difference concerns measurement rather than the underlying
construct.
\end{proposition}

The result follows because every service point has inclusion probability
$n_{im}/N_{im}$ in the control sample. It establishes a common finite-record
target, but not a pure precision effect: differences between nonlinear fitted
shares depend on the mediators' full conditional distributions, not only their
variances.

Matching the precision of the first-set mediator requires more. Suppose the
service points are exchangeable given a latent rate $\pi$, with conditional
variance $v$ and common pairwise covariance $q$, and the number $n$ of
first-set service points is fixed given $\pi$ and does not depend on their
values.

\begin{proposition}[Matched conditional variance]
\label{prop:variance}
Under these conditions, the first-set mediator and a without-replacement
control containing $n$ points have the same conditional error variance:
\[
\operatorname{Var}
\left(
R^{\mathrm{set\,1}}\mid\pi
\right)
=
\operatorname{Var}
\left(
R^{\mathrm{ctrl}}\mid\pi
\right)
=
\frac{v}{n}
+
\left(1-\frac{1}{n}\right)q.
\]
If the points are conditionally independent Bernoulli observations, this
reduces to $\pi(1-\pi)/n$.
\end{proposition}

This is why the control is sampled without replacement: a with-replacement
sample adds variation from repeatedly drawing the observed full-match pool,
making it noisier than the statistic it is meant to match
(Section~\ref{supp-sup:proofs} of the supplementary material gives the
derivation). The conditions of Proposition~\ref{prop:variance} should not be
taken for granted in tennis, since a set ends when one player wins it, making
the number of service points an outcome-dependent stopping time; matching the
observed count does not establish that the two statistics have the same
conditional error distribution. We therefore call the primary construction
point-count-matched. Its contrast with the full-match mediator has a
design-based interpretation conditional on the observed service record, and is
a contrast in measurement precision about a latent rate only under the
exchangeability and stopping assumptions of the proposition, whose failure
modes the reliability analysis and simulation examine.

Measurement error describes the direction in which an imperfect match can
matter. With $\kappa_s$ the conditional reliability of a mediator contrast,
the classical linear attenuation approximation gives
\begin{equation}
\widehat\lambda_s
=
\kappa_s\lambda_s,
\qquad
\widehat c_s
=
c_s+\lambda_sg_s(1-\kappa_s),
\qquad
\kappa_s=
\frac{\operatorname{Var}(\Delta X_{\mathrm{true}}\mid\Delta\theta)}
     {\operatorname{Var}(\Delta X_{\mathrm{obs}}\mid\Delta\theta)}.
\label{eq:eiv}
\end{equation}
The first relation attenuates the estimated mediator coefficient; the second
transfers the omitted part of the pathway to the ability coefficient, so
\[
\widehat c_s+\widehat\lambda_sg_s
=
c_s+\lambda_sg_s,
\]
and the total coefficient is preserved along this attenuation path. For a
linear outcome, lower reliability necessarily reduces the mediated effect and
its share. For the logistic outcome the result is local, because both the
residual scale and the curvature of the link affect probability-scale
estimands, so we evaluate the direction over the fitted posterior rather than
treating it as a general property of logistic mediation models.

When the conditional reliabilities of the first-set and control mediators
differ, a reliability-matched control multiplies the control's point count by
the factor equating the two reliabilities under an inverse-variance relation;
Section~\ref{supp-sup:proofs} of the supplementary material gives the factor
and its derivation. With dependent points the covariance term creates a
variance floor, so the target reliability need not be attainable at any count.
That analysis is therefore a sensitivity to the matching criterion, not a
corrected primary estimate.

The fitted summaries become causal mediation effects only under stronger
conditions: consistency, no interference, positivity, exchangeability for the
exposure and mediator relations, no unmeasured mediator--outcome confounding
after conditioning on ability and baseline variables, no exposure-induced
mediator--outcome confounder, and the required temporal ordering
\citep{robins1992identifiability,pearl2001direct,imai2010general,
vanderweele2015explanation}. The expanding-window construction ensures that
$\theta^-$ uses only earlier seasons. A first-set mediator is earlier than the
final result but still belongs to that outcome window; only the second-set
analysis places it entirely before the modeled outcome. Neither rules out an
unobserved match-level shock, such as injury, fatigue or changing court
conditions, affecting both performance and outcome, which we examine through
sensitivity analysis rather than treat as identified. Latent ability is also a
model-defined exposure rather than a manipulable treatment. Absent those
conditions and a defensible intervention interpretation,
\eqref{eq:de}--\eqref{eq:te} are surface-specific structural pathway
summaries.

Leaving one match out of the ability analysis would not resolve this. It would
keep that match from contributing to its own ability estimate, but gives
neither independence from persistent form, health, scheduling and opponent
selection nor an exclusion restriction, and past results may act on later ones
through seeding, confidence and preparation. Expanding-window estimation
preserves temporal ordering; it does not create an instrument. Proofs, the
stopping-time counterexample and the reliability calculations are in
Sections~\ref{supp-sup:proofs} and~\ref{supp-sup:eiv-details} of the
supplementary material.

\subsection{Posterior computation}
\label{sec:inference}

Both stages are sampled with a purpose-written Hamiltonian Monte Carlo
implementation using analytic gradients
\citep{neal2011mcmc,betancourt2017conceptual} and unconstrained parameters: the
zero-sum restrictions in \eqref{eq:constraints} are imposed through orthonormal
Helmert bases, the surface specialization has a noncentred representation
avoiding the funnel geometry induced by its hierarchical scale, and the
sign-restricted coefficients are sampled on the log scale with the required
Jacobian terms. The step size is adapted by dual
averaging and a diagonal metric is estimated over expanding warmup windows
\citep{hoffman2014nuts}; all adaptation ends with warmup, so the retained
iterations come from a fixed kernel with a Metropolis correction.

Computation follows the modular structure in \eqref{eq:cut}. For each target
season the ability model is fitted to earlier seasons only, and the cut
posterior for the pathway parameters is approximated by refitting the pathway
model conditional on each of $K$ retained ability draws and pooling the
conditional samples. In the point-count-matched arm a different control
realization is paired with each ability draw and standardized within surface
before fitting, so the pooled sample propagates uncertainty in pre-match
ability and in the points drawn to build the mediator. At no stage can the
pathway data update the ability distribution.

Pathway summaries are computed by $g$-computation: residual mediator contrasts
are drawn once and reused across ability interventions,
\eqref{eq:de}--\eqref{eq:te} are evaluated by Monte Carlo, and the mediated
share is formed within each draw rather than as a ratio of posterior means.
Contrasts among constructions use \eqref{eq:comparison}, with the same ability
draw and $g$-computation random numbers in each arm but conditionally
independent pathway draws. The
coupling propagates common ability uncertainty without estimating the residual
covariance induced by reusing the same outcomes; repeated-sampling calibration
is examined in simulation instead. The full parameterization, leapfrog and
adaptation settings, priors and prior predictive checks, gradient verification
and convergence diagnostics are in
Sections~\ref{supp-app:priors}--\ref{supp-sup:diagnostics} of the supplementary
material.

\subsection{Simulation evidence}\label{sec:simulation}

We simulate from each stage to check parameter recovery, and from the point
level to check that the decomposition responds to the two mechanisms it
separates. The ability model recovers the identified match-level predictor,
with fitted log odds correlating 0.968 with the truth at a calibration slope of
0.920, and the performance model recovers its coefficients within Monte Carlo
error. The point-level experiment builds the three mediators as the pipeline
builds them and varies two dials: $\gamma$, the weight of realized points in
the outcome equation, which is outcome overlap, and $\varsigma$, a
segment-level shock giving the opening set its own latent rate, which is
construct change. With both off, the window-construction components cover zero
under both outcomes. Construct change alone produces a positive match-outcome
component with no overlap present. Overlap alone moves the component only when
the outcome follows the first set, because with conditionally independent
points the opening set and the control coincide under the match outcome, a
property of the generator rather than a finding. Over four repeated-sampling
cells the estimator is close to unbiased with conservative 90\% intervals:
where the truth is zero, an interval excludes it in 0--2\% of datasets against
a nominal 10\%. Section~\ref{supp-sup:sim} of the supplementary material gives
the design, the full grid (Table~\ref{supp-tab:simulation}) and the calibration
study (Table~\ref{supp-tab:coverage}).

\section{Real data analysis}\label{sec:results}
\label{sec:results-ability}

This section applies the methodology of Sections~\ref{sec:model}
to eleven seasons of professional tennis. The application serves two purposes.
First, it is a case study in the practical question that motivates the paper:
when an analyst decomposes the effect of pre-match ability into a pathway
through in-match performance, how much of the answer is driven by how the
performance mediator is constructed rather than by the sport itself? Second,
it demonstrates that the proposed control construction, the comparison
distribution in \eqref{eq:comparison}, and the reliability diagnostics can all
be computed from data that national tours, broadcasters, and analytics
departments already collect, and that they yield findings with direct
consequences for how performance statistics are used in the sport.

\subsection{Ability estimates and predictive performance}\label{sec:forecast}

\begin{figure}[htbp]
\centering
\setstretch{1}
\includegraphics[width=0.95\textwidth]{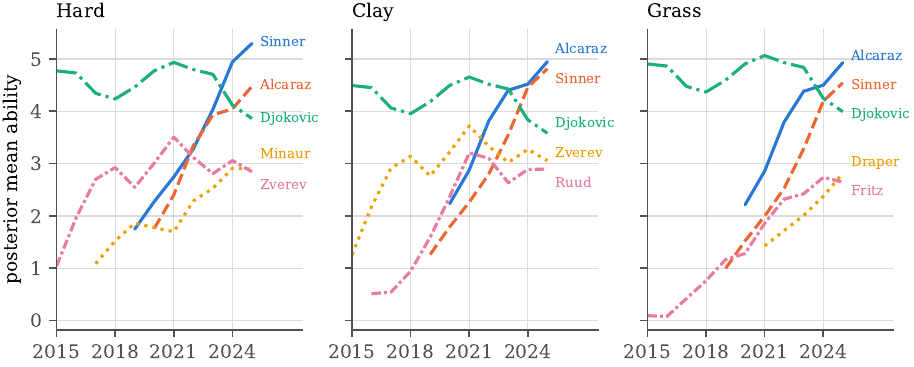}
\caption{Posterior mean ability trajectories from 2015 to 2025 for the five
highest-rated players on each surface at the end of the study period.}
\label{fig:ability-validation}
\end{figure}

Figure~\ref{fig:ability-validation} shows how the leading surface-specific
abilities evolved. The leading players are similar across surfaces but the
ordering is not identical, which quantifies, with posterior uncertainty, how
far ``the best player'' is a surface-dependent statement. Partial pooling is
most visible on grass, where the shorter calendar carries less information. The
five highest posterior means on each surface, with intervals and match counts,
are in Table~\ref{supp-tab:top-abilities} of the supplementary material.

The pathway analysis conditions on pre-match abilities, so those abilities must
be demonstrably competitive as forecasts, though they need not be the best
available forecaster. Scored on the untouched portion of the 2026 season, the
model attains the lowest log loss and Brier score and the highest accuracy
among seven comparators (Table~\ref{supp-tab:validation} of the supplementary
material). Paired per-match $t$-statistics separate it clearly from the frozen
Elo, the surface-blended Elo and the static Bradley--Terry model ($t=3.80$,
$2.62$ and $3.53$), but not from either ranking baseline ($t=1.06$ frozen,
$0.32$ with live ranks) or the prequentially updated Elo ($t=1.24$). The
hold-out also runs only to late spring and contains no grass-court matches, so
the grass ratings are not tested out of sample, and two comparators read the
hold-out season while the model is frozen at the end of 2025;
Section~\ref{supp-sup:forecast} of the supplementary material gives both
features in full. We therefore report this as evidence that the pre-match
abilities are credible, not as a forecasting claim. Its practical reading is
that one hierarchical model, frozen at the season boundary and refit annually,
matches the day-to-day accuracy of ratings updated after every match. The
calibration slope of 0.857 indicates mild overconfidence, consistent with
updating ability between seasons rather than within them;
Figure~\ref{supp-fig:calibration} of the supplementary material shows the
reliability curve.

\label{sec:ppc}
Two posterior predictive checks and an outlier scan appear in full in
Section~\ref{supp-sup:checks} of the supplementary material. The model
reproduces the upset rate, 0.3128 observed against 0.3122 replicated with a
posterior predictive $p$-value of 0.440, and the fitted probabilities are
pulled inward in both tails (Table~\ref{supp-tab:calibration-insample}), the
signature of hierarchical shrinkage. Ranking matches by predictive surprise
returns recognisable upsets, and it surfaced a data issue rather than a
sporting one: several high-surprise results end in retirement, where the
outcome is not produced by the play the mediator measures.
Section~\ref{supp-sup:robustness} of the supplementary material refits the
decomposition without them.

\subsection{Pathway effects and measurement windows}
\label{sec:pathway}
Posterior summaries of the second-stage coefficients are in
Table~\ref{supp-tab:path-coefficients} of the supplementary material. For the
first-set mediator, own ability is positively and opponent ability negatively
associated with service performance, and $|\beta^{\mathrm{opp}}_s|$ is of the
same order as $\beta^{\mathrm{own}}_s$ on every surface, so both players
contribute materially to the mediator. Whether the opponent term is larger is
less clear: the posterior probability that
$|\beta^{\mathrm{opp}}_s|>\beta^{\mathrm{own}}_s$ is 0.83 on hard, 0.87 on clay
and 0.07 on grass, and the priors in Section~\ref{supp-app:priors} of the
supplementary material place $\beta^{\mathrm{own}}$ at a larger location. We
therefore read service performance as a property of the \emph{matchup} rather
than of the server alone: a player's serve statistics are largely a statement
about who that player has faced. That matters whenever they are compared across
players who faced opponents of different quality, and
Section~\ref{sec:industry} develops the implication for leaderboards, scouting
and contract evaluation.

Table~\ref{tab:pathway} compares the three constructions, all fitted to the
same matches with the same pre-match abilities and outcome model and differing
only in the period the mediator covers and the number of service points behind
it.

\begin{table}[htbp]
\centering
\setstretch{1}
\caption{Model-implied changes in win probability for a one-unit standardized
pre-match ability contrast, by mediator. The first two blocks are the comparison
the literature would make; the control block holds the window at the whole match
while reducing the sample size to the first set's. Entries are posterior means
with 90\% credible intervals; the final column is the mediated share of the
total, with its own 90\% interval.}
\label{tab:pathway}
\small
\begin{tabular}{lrrrr}
\toprule
Surface & Direct & Mediated & Total & Mediated share\\
\midrule
\multicolumn{5}{l}{\emph{First set (set-1 window, set-1 sample size)}}\\
\quad Hard & 0.050 [0.040, 0.060] & 0.054 [0.048, 0.060] & 0.104 [0.092, 0.116] & 0.52 [0.47, 0.58]\\
\quad Clay & 0.065 [0.050, 0.080] & 0.044 [0.036, 0.052] & 0.109 [0.093, 0.125] & 0.41 [0.33, 0.48]\\
\quad Grass & 0.055 [0.033, 0.080] & 0.048 [0.037, 0.060] & 0.104 [0.079, 0.129] & 0.47 [0.35, 0.60]\\
\addlinespace
\multicolumn{5}{l}{\emph{Control (full-match window, set-1 sample size)}}\\
\quad Hard & 0.036 [0.027, 0.045] & 0.067 [0.060, 0.074] & 0.103 [0.092, 0.114] & 0.65 [0.59, 0.71]\\
\quad Clay & 0.044 [0.030, 0.058] & 0.061 [0.051, 0.071] & 0.105 [0.089, 0.120] & 0.58 [0.49, 0.68]\\
\quad Grass & 0.047 [0.027, 0.069] & 0.057 [0.042, 0.073] & 0.104 [0.080, 0.128] & 0.55 [0.42, 0.69]\\
\addlinespace
\multicolumn{5}{l}{\emph{Full match (full-match window, full sample size)}}\\
\quad Hard & 0.005 [0.003, 0.008] & 0.101 [0.092, 0.110] & 0.106 [0.096, 0.116] & 0.95 [0.92, 0.97]\\
\quad Clay & 0.010 [0.005, 0.017] & 0.092 [0.081, 0.104] & 0.103 [0.090, 0.115] & 0.90 [0.84, 0.95]\\
\quad Grass & 0.013 [0.006, 0.023] & 0.089 [0.072, 0.107] & 0.103 [0.083, 0.122] & 0.87 [0.79, 0.94]\\
\bottomrule
\end{tabular}
\end{table}

For the temporally ordered first-set mediator the two pathways are of similar
size: on hard courts a one-standard-deviation ability advantage raises the win
probability by 0.104, of which 0.054 passes through first-set service
performance and 0.050 remains direct, a mediated share of 0.52. The full-match
mediator gives a share of 0.95 and a direct effect of 0.005. Since the same
service points enter the mediator and the result, that reads more naturally as
outcome overlap than as evidence that serving is the only route from ability to
winning. The main empirical finding is therefore negative: the near-complete
mediation a full-match analysis reports, which a practitioner would read as
``winning is almost entirely serving'', reflects how the mediator is
constructed rather than a property of tennis.

\label{sec:headline}
The control arm splits the conventional gap into a point-count and a
window-construction component.

\begin{table}[htbp]
\centering
\setstretch{1}
\caption{Decomposition of the gap between the first-set and full-match mediators
into a window-construction component (first set $\to$ control, the window widens at fixed
sample size) and a point-count component (control $\to$ full match, the sample size
grows at fixed window), with the outcome being the match. The three share
columns are posterior means; the two component columns are posterior means with
90\% credible intervals, all on the mediated-share scale. The final column is
the window-construction component as a share of the two-arm gap, formed from
posterior means; it is not an overlap share because it combines
overlap with the change in the measured construct.}
\label{tab:decomposition}
\small
\begin{tabular}{lrrrrrr}
\toprule
Surface & \multicolumn{3}{c}{Mediated share} & Window & Point count & Window \%\\ \cmidrule(lr){2-4}\cmidrule(lr){5-6} & set 1 & control & full & component & component & of gap\\
\midrule
Hard & 0.522 & 0.649 & 0.950 & 0.127 [0.053, 0.203] & 0.301 [0.236, 0.364] & 30\%\\
Clay & 0.405 & 0.582 & 0.899 & 0.177 [0.066, 0.288] & 0.317 [0.213, 0.420] & 36\%\\
Grass & 0.473 & 0.553 & 0.872 & 0.080 [-0.090, 0.247] & 0.319 [0.173, 0.464] & 20\%\\
\bottomrule
\end{tabular}
\end{table}

The control spans the whole match, so it carries whatever outcome overlap the
conventional mediator carries, but is computed from the opening set's number of
service points; under Proposition~\ref{prop:variance} it also has the first-set
mediator's conditional error variance. On hard courts its mediated share is
0.65, between the other two, with a window-construction component of 0.127
[0.053, 0.203] and a point-count component of 0.301 [0.236, 0.364]. The
window-construction component is the smaller part of what the conventional
comparison reports, between 20\% and 36\% of the gap depending on surface, so
most of the apparent difference between an early-window and a full-match
analysis is the arithmetic of counting more points. How much of it is outcome
overlap this design does not say, since the component combines overlap with a
change in the measured construct. Clay shows the largest component and grass
the least precise estimate, as the sample sizes lead one to expect.

Intervals for these contrasts use the comparison distribution in
\eqref{eq:comparison}, so they quantify uncertainty under that stated coupling
rather than estimating a covariance identified by a joint model for the three
mediators. Repeated-sampling behavior under the simulation generator and the
pooling construction are in Sections~\ref{supp-sup:simfail}
and~\ref{supp-sup:pooling} of the supplementary material.

Conditional on the complete point record the control is an unbiased subsample
of the raw full-match proportion with binomial sampling variance close to the
first-set statistic's, which supports the point-count comparison without
identifying a pure measurement-error effect. We therefore report both
quantities as construction contrasts.

The window-construction component is smaller than the overlap intuition
suggests, for a reason that transfers beyond tennis. The outcome is who won the
match and the first set is part of that match, so a first-set mediator is
earlier than the end of the outcome but not outside it: its points are among
those that decide the thing being explained. Widening the window moves the
mediator from one region inside the outcome to a larger region inside the same
outcome, adding overlap to a baseline that was never overlap-free. Temporal
ordering and freedom from outcome overlap are therefore distinct properties of
a mediator, and an analysis can have the first without the second.

\subsection{Temporal ordering, robustness, and replication}
\label{sec:strict}

Because a first-set mediator is observed earlier than the match result but
remains part of it, we repeat the decomposition with the outcome defined as
winning the second set. The mediator is then entirely prior to the modeled
outcome, while the control still samples the full match, including second-set
points. This is a positive-control analysis for outcome overlap, not an
estimate of how much overlap occurs naturally.

The window-construction component grows when the outcome moves outside the
first-set window. On hard courts it rises from 0.127 $[0.053,0.203]$ for the
match outcome to 0.311 $[0.229,0.396]$ for the second-set outcome, about
two-and-a-half-fold, and the clay-court estimate is 0.328 $[0.217,0.442]$. The
pattern is consistent with outcome overlap, though it does not identify overlap
separately from changes in performance across sets. It shows how far the
placement of the outcome window can move a mediated share.

The conclusion is similar after splitting matches by tournament round,
excluding retirements and walkovers, replacing the expanding-window exposure
with a retrospective ability estimate, and perturbing the mediator--outcome
association with an unmeasured match-level shock; permuting outcomes within
surface removes both components. These checks cannot remove selection into the
charted sample or establish the causal assumptions of
Section~\ref{sec:estimands}; Section~\ref{supp-sup:robustness} of the
supplementary material gives every estimate.

The analysis was repeated independently for the WTA tour with 26,591 ability
matches and 2,332 linked pathway matches. Against the match outcome the
hard-court window-construction component is 0.109 $[-0.004,0.223]$, against
0.127 $[0.053,0.203]$ for the ATP sample; against the second-set outcome it
rises to 0.316 $[0.192,0.436]$. The replication reproduces both the direction
of the decomposition and its dependence on whether the mediator lies inside the
modeled outcome window. The tours differ in playing style, calendar and match
format at the largest events, so the agreement suggests these construction
effects belong to the analytic design rather than to one tour.

Historical ability is thus associated with subsequent service performance, and
both with match results. The much larger mediated share from full-match
performance is not evidence that serving is the only route from ability to
winning; it reflects a mediator measured with more points and over a period
overlapping the outcome, and the control construction diagnoses both changes
without assigning a causal share to either.

\subsection{Reliability analysis without point-level data}
\label{sec:eiv}

When only an aggregate performance statistic and its denominator are available,
the errors-in-variables approximation in \eqref{eq:eiv} is a diagnostic rather
than a generally valid correction. We estimate the sampling variance by
$p(1-p)/n$ and compare it with the residual mediator variance
$2\sigma_X^2(1-\rho)$, which assumes conditionally independent Bernoulli points
and a fixed count, though tennis points are dependent and the number played is
outcome-dependent.

Table~\ref{supp-tab:kappa} of the supplementary material shows why equal point
counts should not be called equal reliability, which matters because matching
on sample size is the default fairness criterion in applied work. The control
has nearly the first-set mediator's plug-in sampling variance but lower
conditional reliability on every surface, because the two constructions differ
in signal variation. Equal $n$ does not imply equal information, so the
point-count component is not an identified measurement-error effect. Matching
reliability instead moves the hard-court window-construction component from
0.127 to 0.215 and the clay-court component from 0.177 to 0.329, which we
report as sensitivity to the matching criterion, not as a corrected
decomposition.

The regression-calibration diagnostic is
\begin{equation}
D_s=\widehat c_s-\frac{\widehat\lambda_s}{\kappa_s}g_s(1-\kappa_s),
\label{eq:diagnostic}
\end{equation}
the implied direct coefficient after attenuation correction. It is negative in
10 of the 12 mediator-by-surface cells, so the approximation is incompatible
with the positive-direct-effect restriction in most fitted configurations. An
unrestricted hard-court refit reinforces the warning: the full-match direct
coefficient is $-0.091$ $[-0.175,-0.009]$ and the mediated share is 1.07, while
the first-set estimate is essentially unchanged. A share above one signals that
the restricted model is at its boundary, so the near-boundary full-match share
is sensitive to the restriction and prior rather than evidence of complete
mediation.

Section~\ref{supp-sup:eiv-details} of the supplementary material gives the
reliability derivation, correction validation, unrestricted fits,
reliability-matched decompositions and fixed-point calculations. Across all
reported configurations the maximum rank-normalized split-$\widehat R$ is
1.012, the minimum bulk effective sample size is 720, and no divergent
transitions occur among retained draws
(Section~\ref{supp-sup:diagnostics} of the supplementary material).

\subsection{Implications for the sports industry}
\label{sec:industry}

Four consequences follow from the results rather than from sporting intuition.
First, because $|\beta^{\mathrm{opp}}_s|$ is of the same order as
$\beta^{\mathrm{own}}_s$ (Section~\ref{sec:pathway}), a serve percentage is
partly a statement about the opponent, so leaderboards quoting it without
opponent adjustment let schedule imbalance confound any comparison or
performance clause built on them; the fitted coefficients supply the
adjustment, with posterior uncertainty, from abilities the model already
produces. Second, ranking matches by predictive surprise
(Section~\ref{sec:ppc}) screens a data-quality or integrity pipeline, and here
it surfaced retirements, which are censoring events rather than sporting
anomalies. Third, the decomposition of Section~\ref{sec:headline} cautions any
mediation-style analysis whose mediator is measured on a window overlapping the
outcome: the mediated share rose by a factor of 1.6 to 2.2 between the
first-set and full-match constructions across both tours and all surfaces, and
matching on sample size does not fix it, since equal counts are not equal
reliability (Table~\ref{supp-tab:kappa}). The diagnostic needs only the same
data, re-windowed, and the ATP--WTA agreement suggests it transfers to other
sports recorded as sequences of within-contest events. Fourth, the hold-out
comparison of Section~\ref{sec:forecast} shows a pooled rating frozen at the
season boundary to be indistinguishable in accuracy from ratings updated after
every match, which matters where continuous updating is an operational
liability.

\section{Discussion}\label{sec:discussion}

When a mediator is measured cannot be separated from how precisely it is
measured. Full-match service performance rests on more points than first-set
performance, so comparing the two confounds timing with precision. The
point-count-matched control pulls them apart: against the full-match mediator
it isolates a point-count component, and against the first-set mediator it
holds the count fixed and isolates the remaining window-construction
component. Reading the first contrast as pure measurement
precision requires the assumptions of Section~\ref{sec:estimands}. In our data
the point-count component is the larger, but the window component still matters
and its meaning is ambiguous, mixing mediator--outcome overlap with real
changes in performance across sets. The second-set analysis makes that concrete:
a mediator recorded earlier in a match is not necessarily measured before the
outcome. Timing and precision are distinct design choices, and both shape the
pathway an analyst will attribute to performance.

These conclusions come with limits. The Match Charting Project is a volunteer
archive that over-covers prominent players and marquee matches, so the linked
sample is not representative of professional tennis, and cross-round
comparisons cannot rule out selection because every match comes from the same
archive. The simulation reproduces the expected behavior of the decomposition
but simplifies outcome-dependent match length, dependence among points and the
imputation procedure. Unmeasured match-level shocks remain possible; the
sensitivity analysis probes them but the data do not identify them. The model
holds surface specialization fixed and lets ability move only between seasons.
Finally, the modular procedure targets a cut posterior, which protects the
pre-match meaning of ability but whose frequentist calibration need not match
a joint posterior's.

Point-level records would let mediators be built over progressively longer
stretches of a match, tracing the pathway as the window widens; serving,
returning and pressure-point performance could be separate mediators; ability
and specialization could evolve within a season. More broadly, studies using
temporally aggregated mediators should treat their construction as part of the
inferential design: report the number of contributing observations, compare
windows at a common precision, and measure the mediator outside the outcome
window when a temporal interpretation is intended. Without these safeguards, a
change in the mediated effect may come from measurement error, temporal
overlap, a shifting construct, or all three at once.



\section{Data Availability Statement}\label{data-availability-statement}

The match results and point-level charting records analysed here are publicly
available from the archives maintained by Jeff Sackmann under a CC BY-NC-SA
4.0 licence \citep{sackmann_atp,sackmann_wta,sackmann_mcp}. The code that
downloads the data, records their provenance, fits both stages and generates
every reported table, figure and value accompanies the supplementary
material.

\phantomsection\label{supplementary-material}
\medskip

\begin{center}
{\large\bf SUPPLEMENTARY MATERIAL}
\end{center}

\begin{description}[leftmargin=0pt,itemsep=1pt,parsep=0pt,topsep=2pt]
\item[Supplementary material:] Sections S1--S14: data linkage, proofs, the
control mediator, priors, posterior computation and diagnostics, simulation
details, model checking, the forecast comparison, additional empirical results,
interval construction, robustness and sensitivity checks, and the reliability
analysis. 
\item[Reproduction archive:] Python code for data preparation, both model
stages, the simulations and every table and figure; the fitted result archives;
and a provenance manifest with the URL, retrieval date and SHA-256 digest of
every downloaded file. (zip)
\end{description}

\begin{spacing}{1.2}
\bibliography{Bibliography-MM-MC}
\end{spacing}

\end{document}